\newcommand{\vub}{\mbox{$|V_{ub}|$}}

\newcommand{\dtopilnu}{\mbox{$D^0 \to \pi^- \ell^+ \nu$}} 
\newcommand{\btopilnu}{\mbox{$B^0 \to \pi^- \ell^+ \nu$}} 
 
\newcommand{\dtopienu}{\mbox{$D^0 \to \pi^- e^+ \nu$}} 
\newcommand{\dtopimunu}{\mbox{$D^0 \to \pi^- \mu^+ \nu$}} 
\newcommand{\dtoklnu}{\mbox{$D^0 \to K^- \ell^+ \nu$}}
\newcommand{\dtokenu}{\mbox{$D^0 \to K^- e^+ \nu$}}
\newcommand{\dtokmunu}{\mbox{$D^0 \to K^- \mu^+ \nu$}}
\newcommand{\dtokslnu}{\mbox{$D^0 \to K^{*-} \ell^+ \nu$}}

\newcommand{\mhlnu}{\mbox{$m_{hl\nu}$}}

\newcommand{\pmiss}{\mbox{${\vec{p}}_{\rm miss} $}}

\newcommand{\qs}{\mbox{$q^2$}}
\newcommand{\Rz}{\mbox{$R_0$}}
\newcommand{\mpole}{\mbox{$m_{\rm pole}$}}
\newcommand{\rfo}{ \mbox{${|f^{\pi}_{+}(0)| / |f^K_{+}(0)|}$}}

\newcommand{\gev}{\mbox{{\rm GeV}}}
\newcommand{\mev}{\mbox{{\rm MeV}}}

\documentclass{ws-ijmpa}

\begin{document}

\markboth{F. Liu}
{Precision Measurements of the  Semileptonic Charm Decays ...}  

%
\catchline{}{}{}{}{}
%

\title{Precision Measurements of the  Semileptonic Charm Decays 
{\boldmath$D^0 \to \pi^- \ell^+ \nu$} and {\boldmath$D^0 \to K^- \ell^+ \nu$} }

\author{\footnotesize F. Liu \\ Representing the CLEO Collaboration}

\address{Southern Methodist U. Group, Wilson Lab, Cornell University, Ithaca, New York 14853}

\maketitle


\begin{abstract} 
We investigate the decays $D^0 \to \pi^- \ell^+ \nu$ and $D^0 \to K^- \ell^+ \nu$, where $\ell$ is $e$ or $\mu$, using approximately 7 ${\rm fb}^{-1}$ of data collected with the CLEO III detector.  We find $\Rz \equiv {\cal B}(\dtopienu)/{\cal B}(\dtokenu) = 0.082 \pm 0.006 \pm 0.005$.  Fits to the kinematic distributions of the data provide parameters describing the form factor of each mode.  Combining the form factor results and \Rz\ gives $|f^{\pi}_{+}(0)|^2 |V_{cd}|^2/|f^K_{+}(0)|^2 |V_{cs}|^2 =   0.038^{+0.006+0.005}_{-0.007-0.003}$.
\end{abstract}


\section{Introduction}
The quark mixing parameters are fundamental constants of the weak interaction.  Measuring them also tests the unitarity of the quark mixing (CKM) matrix.
Semileptonic decays have provided most quark coupling data~\cite{Motiv}.  

 We present a study of the decays \dtopilnu\ and \dtoklnu, where $\ell = e\ {\rm or}\ \mu$.  We measure the ratio of their branching fractions, $\Rz \equiv {\cal B}(\dtopienu)/{\cal B}(\dtokenu)$, and, for the first time for \dtopilnu, parameters describing the form factors.  The study of the \dtopilnu\ form factor is particularly interesting because it tests predictions for that of the closely related decay \btopilnu, which provides \vub. Charge conjugate modes are implied throughout this paper. 
This paper is an abbreviated version of Ref.~\cite{prl3}.

The differential partial widths for \dtopilnu\ and \dtoklnu 
~with the lepton masses neglected, in terms of the form factor $f_+(q^2)$, are
$$
\frac{d\Gamma}{d\qs}(D\to h \ell\nu) = \frac{G_F^2}{24\pi^3}p_{h}^3|V_{cd(s)}|^2  |f_+^{h}(q^2)|^2  .
$$
Here $h$ is $\pi$ or $K$, and \qs\ is the invariant mass squared of the lepton-neutrino system with $m_\ell^2\le q^2\le m_D^2-m_h^2$, $m_D$ and $m_h$ are the $D$ meson
and hadron masses, respectively.  To reduce the form factor sensitivity of \Rz\ and determine the \qs\ distributions, the yields are extracted in bins of \qs.

\section{Analysis Techniques}
We use $e^+ e^- \to c \bar{c}$ events collected at and just below the $\Upsilon(4S)$ resonance with the CLEO III detector~\cite{CLEOIII}.  
A major challenge for this analysis is the contamination of the \dtopilnu\ sample by \dtoklnu\ decays, which are about a factor of 10 more common.  The use of a Ring-Imaging Cherenkov detector (RICH) and specific ionization in the drift chamber ($dE/dx$) reduces this contamination dramatically by distinguishing $K$ from $\pi$ mesons.  The resulting efficiency and misidentification probability suppress misidentified \dtoklnu\ decays to 15\% of the \dtopilnu\ signal.

$D^0$ candidates are reconstructed from identified lepton, hadron ($\pi$ or $K$), and neutrino combinations.  
The missing momentum of the event (\pmiss) provides the first estimate of the neutrino momentum: $\vec{p}_\nu=-\pmiss$. To improve the neutrino momentum resolution, we further impose the mass constraint $\mhlnu=m_{D^0}$ (we use PDG ~\cite{PDG} masses throughout).  This constraint improves the neutrino momentum resolution by a factor 
of two. The detail of event selection can be found in Ref.~\cite{prl3}.

We require that all $D^0$ candidates come from the decay $D^{*+} \to D^0 \pi^+$.
We reconstruct the $D^{*+}$ by pairing a pion with the appropriate charge (the ``soft'' pion, $\pi_{\rm s}$) with the $D^0$ candidate and then compute the mass difference between the $D^{*+}$ and the $D^0$ candidates, $\Delta m = m_{h\ell\nu\pi_{\rm s}} - \mhlnu$.  
The signal peaks in the region $\Delta m < 0.16$ \gev\ (the ``signal region'') with a root-mean-square width of about 10 \mev.  We use the $\Delta m$ distribution to extract the yields. 


About half the background in the signal region is composed of candidates in which the $\pi_{\rm s}^+$ comes from a $D^{*+}$ decay.  This background is troublesome because it peaks in $\Delta m$.  A Monte Carlo simulation~\cite{geant} shows that the dominant ``peaking'' background~\cite{prl3} in the \dtopilnu\ sample comes from \dtoklnu\ decays in which the $K$ is mis-identified as  a $\pi$, and from candidates in which a lepton from $D^0$ semileptonic decays is paired with a random pion.  The remaining half of the background does not peak because the $\pi_{\rm s}$ is not from a $D^*$ decay (the ``false-$\pi_{\rm s}$'' background).  For the more common \dtoklnu\ mode, the dominant peaking background~\cite{prl3} comes primarily from the decay \dtokslnu 
~with  $K^{*-}\to K^-\pi^0$.


We divide the data into three \qs\ bins: [0, 0.75] (bin 1), [0.75, 1.5] (bin 2), and $>1.5$ \gev$^2$ (bin 3).  
The yield in each \qs\ bin for each of the modes, \dtokenu, \dtokmunu, \dtopienu, and \dtopimunu, is determined from a fit to the $\Delta m$ distribution.  The Monte Carlo simulation~\cite{geant} provides the $\Delta m$ distributions of the signal and backgrounds.  The \dtoklnu\ samples are fit first.  The two free parameters in these fits are the normalizations of the \dtoklnu\ simulated signal and of the false-$\pi_{\rm s}$ background relative to the data.  

\begin{figure}[hbtp]
\begin{center}
\includegraphics*[width=0.9\textwidth]{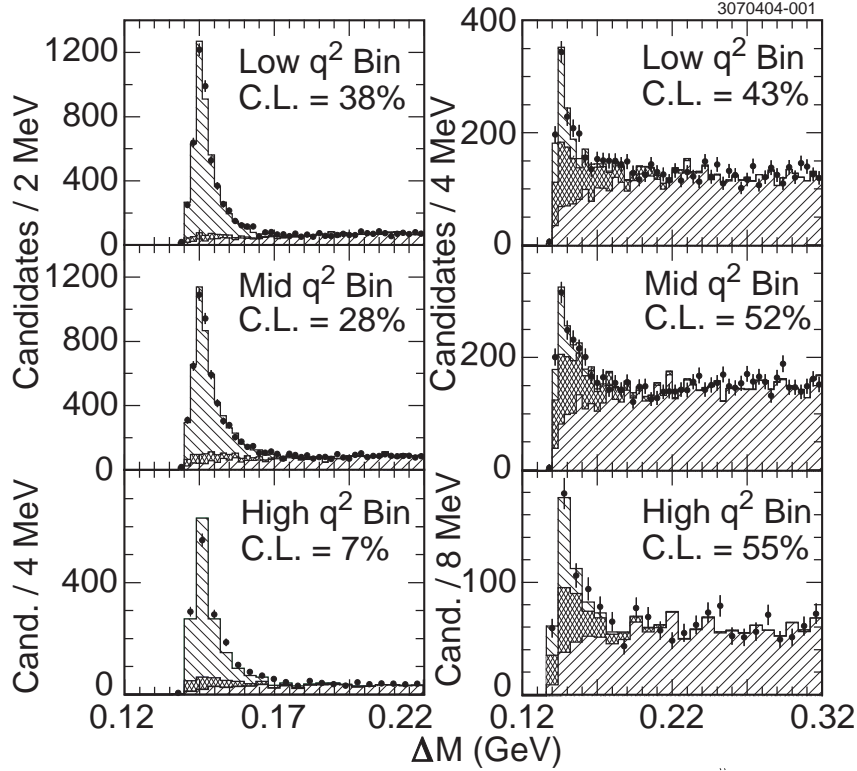} 
\caption{The fits to the $\Delta m$ distributions for \dtokenu\ (left) and \dtopienu\ (right) and their confidence levels (C.L.).  The data (points) are superimposed on the sum of the normalized simulated signal (peaked histogram), peaking background (dark histogram) and false-$\pi_{\rm s}$ background (broad histogram).}
\label{fig:fits}
\end{center}
\end{figure}

Then the  \dtopilnu\ samples are fit.   The normalization of \dtoklnu\ from the \dtoklnu\ fits sets the normalization of the peaking background in the \dtopilnu\ fits.  The two free parameters in these fits are the normalizations of the \dtopilnu\ signal and of the false-$\pi_{\rm s}$ background.   The electron mode fits and their confidence levels are shown in Figure~\ref{fig:fits}.  
The muon fits are similar, but with smaller sample sizes because of the muon momentum and angular restrictions.  


\section{Results and Discussions}
\begin{figure}[htbp]
\begin{center}
\includegraphics*[width=0.9\textwidth]{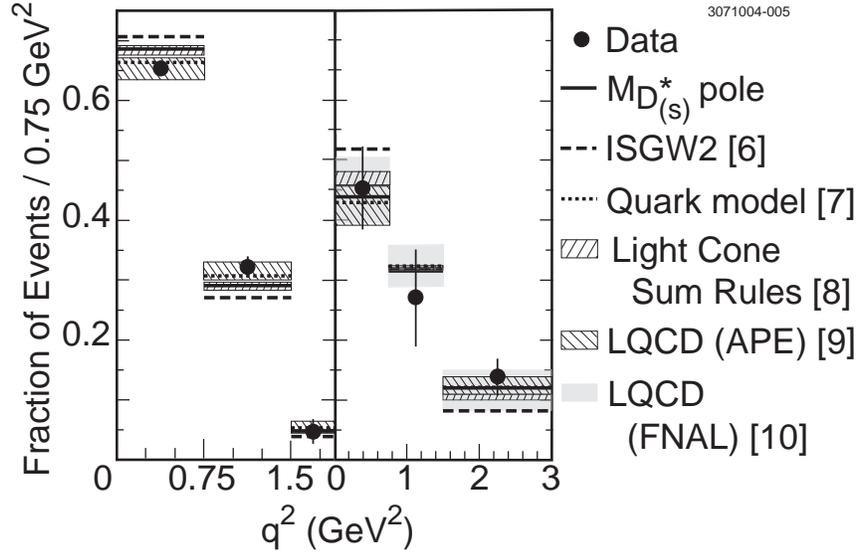} 
\caption{Distributions in \qs\ for \dtoklnu\ (left) and \dtopilnu\ (right), after correcting 
for reconstruction efficiency and smearing in \qs, 
and predictions~$^{[6-10]}$.  
The data include statistical and systematic uncertainties. }
\label{fig:qsqrDist}
\end{center}
\end{figure}

We sum the efficiency corrected yields over \qs\ bins to find $R_{0_e} = 0.085 \pm 0.006 \pm 0.006$ and $R_{0_\mu}= 0.074 \pm 0.012 \pm 0.006$ for the electron and muon modes, respectively, where the first uncertainty is statistical, and the second is systematic. We then compute the normalized \qs\ distributions, which are defined as the fraction of the total corrected yield in each \qs\ bin (since the $D^*$ production rate is undetermined).  They are shown in Figure~\ref{fig:qsqrDist}.  The results combine the electron and muon modes after correcting the muon modes for their reduced phase space.    The systematic uncertainties are dominated by uncertainties in the backgrounds,  inaccuracies in the simulation of the neutrino momentum  reconstruction, 
and hadron misidentification.

%

Combining $R_{0_e}$ and $R_{0_\mu}$ after applying a $+1$\% correction to $R_{0_\mu}$ to account for the reduced muon phase space, gives
$$
R_0 = 0.082 \pm 0.006 \pm 0.005.
$$
This result is consistent with the previous world average 
~\cite{PDG,BES}, but is more precise.

We next determine parameters describing the form factors by fitting the corrected \qs\ distributions.  We first use a simple pole parameterization,
$$
f_+^{h}(q^2) = \frac{f_+^{h}(0)}{1-q^2/m_{\rm pole}^2},
$$
and vary the value of \mpole, constraining the integral over \qs\ to unity.  The quality of the fits is good.  Dominance by a single pole would imply $m_{\rm pole}^{D\to h} = m_{D_{(s)}^*}$.  We find $m_{\rm pole}^{D\to \pi}= 1.86^{+0.10+0.07}_{-0.06-0.03}$ \gev\ and $m_{\rm pole}^{D \to K} = 1.89\pm 0.05^{+0.04}_{-0.03}$  \gev, where the uncertainties are statistical and systematic. 
 We also fit the data with a modified pole distribution~\cite{BK}, 
$$
f_+^{h}(q^2) = \frac{f_+^{h}(0)}{(1-q^2/m^2_{D^*_{(s)}})(1-\alpha q^2/m^2_{D_{(s)}^*})},
$$
to obtain the parameter $\alpha$.  We find $\alpha^{D\to \pi} = 0.37^{+0.20}_{-0.31}\pm0.15$
 and $\alpha^{D\to K} = 0.36\pm 0.10^{+0.03}_{-0.07}$.  
Our results for $m_{\rm pole}^{D \to K}$ and $\alpha^{D\to K}$ suggest the existence of contributions beyond the pure $D_s^*$ pole to the \dtoklnu\ form factor.  For \dtopilnu,  $m_{\rm pole}^{D\to \pi}$ is consistent with the $D^*$ mass, though the precision is sufficient to rule out only large additional contributions. 

Several predictions~\cite{isgw2,melikhov,LCSR2000,LQCDAbada,LQCDElKhadra}
for the form factors are superimposed on our data in Figure~\ref{fig:qsqrDist}.  Most are in satisfactory agreement with the data.  The ISGW2 model~\cite{isgw2}, however, predicts a \qs\ distribution for \dtoklnu\ that peaks lower than the data, and accordingly the $\chi^2$ with our data is poor (18 for 2 degrees of freedom). 

Using the value of \Rz\ and parameterizing the form factors with the results of the modified pole fit, we find 
$$
{|f^{\pi}_{+}(0)|^2 |V_{cd}|^2\over |f^K_{+}(0)|^2 |V_{cs}|^2} = 0.038^{+0.006+0.005}_{-0.007-0.003},
$$
where the uncertainties are statistical ($\pm 0.003$ from \Rz\ and $\pm0.006$ from $\alpha$) and systematic ($\pm 0.002$ from \Rz\ and $^{+0.004}_{-0.002}$ from $\alpha$).  The result is the same within 1\% if we use the simple pole form factor instead.   Using $|V_{cd}/V_{cs}|^2=0.052 \pm 0.001$~\cite{PDG}  gives $\rfo = 0.86\pm0.07^{+0.06}_{-0.04}\pm 0.01$, where the first error is statistical, the second is systematic, and the third is from the CKM matrix elements.  This value is consistent with most expectations for SU(3) symmetry breaking~\cite{melikhov,LCSR2000,LQCDAbada,LQCD2004}.

\section{Summary}
We have presented a new measurement of the ratio of \dtopilnu\ to \dtoklnu\ decay rates.  This result is more precise than 
any previous measurement by a factor of two~\cite{PDG,BES}.   Our data also provide new information on the \dtoklnu\ form factor, a first determination of the \qs\ dependence of the \dtopilnu\ form factor and the first model independent constraint on $|f^{\pi}_{+}(0)| |V_{cd}|/ |f^K_{+}(0)| |V_{cs}|$.  Together, these offer new checks of SU(3) symmetry breaking and the form factors predicted for the semileptonic decays of heavy mesons into light ones.


\end{document}